# Two-dimensional type-II Dirac fermions in a $LaAlO_3/LaNiO_3/LaAlO_3$ quantum well

L. L. Tao* and Evgeny Y. Tsymbal†

*Department of Physics and Astronomy & Nebraska Center for Materials and Nanoscience*
*University of Nebraska, Lincoln, Nebraska 68588, USA*

The type-II Dirac fermions that are characterized by a tilted Dirac cone and anisotropic magneto-transport properties have been recently proposed theoretically and confirmed experimentally. Here, we predict the emergence of two-dimensional type-II Dirac fermions in $LaAlO_3/LaNiO_3/LaAlO_3$ quantum-well structures. Using first-principles calculations and model analyses, we show that the Dirac points are formed at the crossing between the $d_{x^2-y^2}$ and $d_{z^2}$ bands protected by the mirror symmetry. The energy position of the Dirac points can be tuned to appear at the Fermi energy by changing the quantum-well width. For the quantum-well structure with a two-unit-cell-thick $LaNiO_3$ layer, we predict the coexistence of the type-II Dirac point and the closed nodal line. The results are analyzed and interpreted using a tight-binding model and symmetry arguments. Our findings offer a practical way to realize the 2D type-II Dirac fermions in oxide heterostructures.

***Introduction.*** The recent proposal of type-II Weyl fermions[1] has inspired intensive investigations of the counterpart type-II Dirac fermions[2,3,4,5]. The type-II Weyl/Dirac fermions merge at the boundary between electron and hole pockets and exhibit specific anisotropic magneto-transport behavior qualitatively different from the type-I Dirac fermions[1,6,7]. To date, a handful of candidates hosting the three-dimensional (3D) type-II Dirac fermions have been theoretically proposed, such as materials deriving from $PtSe_2$[2], $YPd_2Sn$[3], $VAl_3$[4], and KMnBi[5] families. The experimental evidence of the 3D type-II Dirac fermions in bulk $PtSe_2$[8,9], $PtTe_2$[10], and $PdTe_2$[11] have been reported based on angle-resolved photoemission spectroscopy measurements.

On the other hand, towards the devices miniaturization, it would be beneficial to realize type-II Dirac fermions in two-dimensional (2D) system by analogy to the well-known type-I Dirac fermions in graphene[12]. For example, the existence of the 2D type-II Dirac fermions have been predicted in monolayer $WTe_2$[13], graphene with nitrogen line defects[14], quinoid-type α-$(BEDT\text{-}TTF)_2I$[15]. However, the properties of these free-standing 2D materials systems may change when they are deposited on a substrate for performing experimental measurements. For example, the Dirac cone in a free-standing silicene is destroyed due to the hybridization with an Ag(111) substrate[16]. In this regard, it would be desirable to find 2D systems, which can be accurately controlled and not effected by the conditions of measurements. From this point of view, complex oxide heterostructures may serve as a fertile playground to realize the 2D type-II Dirac fermions in practice. Due to advances in thin-film deposition and characterization techniques, the layered oxide heterostructures can be synthesized with atomic-scale precision and exhibit variety of electronic phenomena not found in bulk constituents[17]. Very recently, 2D type-II Dirac fermions have been observed in a layered oxide $La_{2-x}Sr_xCuO_4$.[18]

Here, we propose the realization of 2D type-II Dirac fermions in the experimental feasible oxide quantum-well structures $LaAlO_3/LaNiO_3/LaAlO_3(001)$[19]. Bulk $LaNiO_3$ (LNO) is a paramagnetic metal[20,21], which can be used as an oxide electrode[22,23], while bulk $LaAlO_3$ (LAO) is a wide-gap insulator[24]. Previous work on LNO/LAO (001) heterostructures has been focused on the orbital engineering to produce novel electronic properties[25-28]. Here, we exploit the effect of quantum confinement of the LNO layer to form the emerging electronic states. We predict that the type-II Dirac points (DPs) appear along the diagonal axes of the 2D Brillouin zone (BZ) and the band dispersion around the DP is strongly anisotropic in the momentum space. Given the experimentally feasible structure with tunable degrees of freedom, the proposed LAO/LNO/LAO quantum-well system represents a promising candidate to realize the 2D type-II Dirac fermions in practice.

***Results and discussion.*** First, we investigate the atomic and electronic structure of a LAO/$(LNO)_n$/LAO quantum well with LNO thickness $n$ of 1.0 unit cell (u.c), as depicted in Fig. 1(a). Density-functional theory (DFT) calculations are performed using the plane-wave ultrasoft pseudopotential method[29] implemented in Quantum-ESPRESSO[30,31,32,33] (further details can be found in the Supplemental Material[34]). Fig. 1(b) shows the calculated relative metal-oxygen (M-O) displacements across the heterostructure. Overall, we see that the displacements are quite small with the slight enhancement ~0.03 Å at the interfaces. This stems from no electrostatic mismatch across the interface, due to the equal formal valence of the $(NiO_2)^-$ and $(AlO_2)^-$ atomic layers. This is contrast to the well-known $SrTiO_3/LaAlO_3$ (001) system where a large La-O displacement at the interface ~0.2 Å is produced between the charged $(LaO)^+$ and neutral $(TiO_2)^0$ atomic layers[35]. Fig. 1(c) shows the local density of states (LDOS) across the quantum well. It is seen that the LDOS at the Fermi energy $E_F$ is non-zero only within the LNO layer, suggesting a nearly perfect 2D electron gas. Examining the LDOS of the LNO layer suggests that the Ni-$t_{2g}$ ($d_{xy}$, $d_{yz}$, $d_{zx}$) orbitals are well below $E_F$ and fully occupied, whereas the Ni-$e_g$ ($d_{x^2-y^2}$, $d_{z^2}$) orbitals are partially occupied and thus determine the electronic structure around $E_F$.

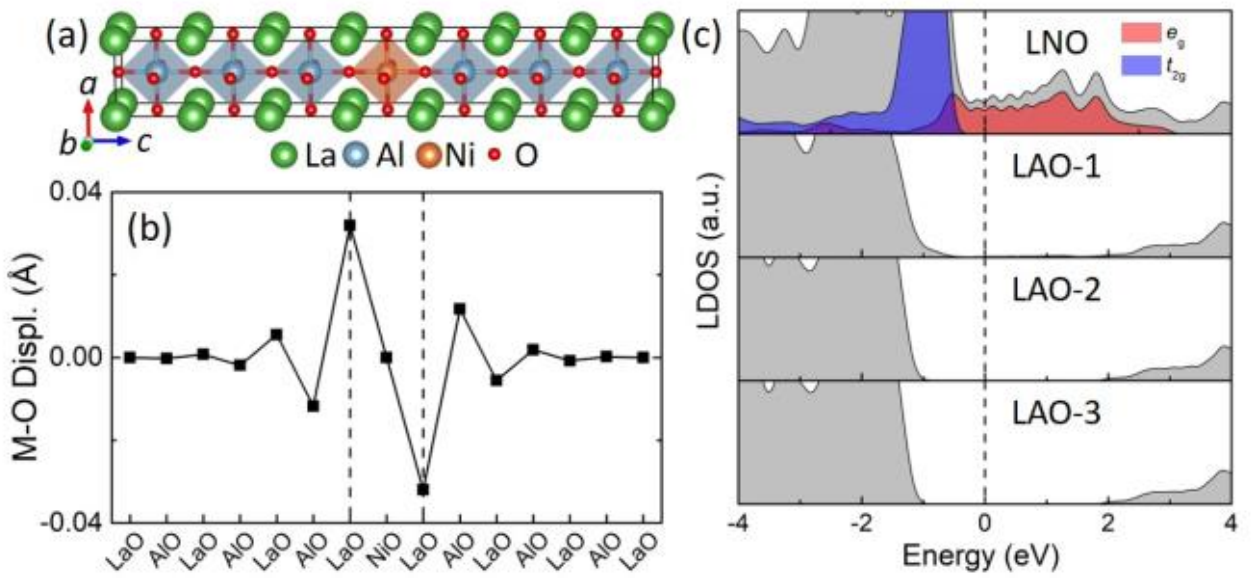

FIG. 1. (a) Atomic structure of the LAO/(LNO)$_1$/LAO. Here (*a*, *b*, *c*) axes is concordant with the (*x*, *y*, *z*) axes. (b) The relative metal-O (M-O) displacements. The two vertical dashed lines denote the interfaces. (c) Local density of states (LDOS) on layers from LNO to deep LAO-3. The vertical dashed line denotes the Fermi energy.

Next, we discuss the electronic structure of the LAO/(LNO)$_1$/LAO quantum well. Fig. 2(a) shows the calculated band structure without spin-orbit coupling (SOC). It is seen that the $d_{x^2-y^2}$ and $d_{z^2}$ bands cross each other along Γ-M. Zooming in around the crossing point (~0.8 eV above $E_F$) reveals a linear dispersion, which is strongly anisotropic in *k* space [Figs. 2(b) and 2(c)]. The band crossing is tilted along Γ-M [Fig. 2(b)], while it is straight perpendicular to Γ-M [Fig. 2(c)]. Fig. 2(d) shows the 3D band structure around the crossing point. It is seen that cutting the bands by the isoenergy plane opens electron and hole pockets touching at the crossing point [Fig. 2(e)]. Due to space inversion symmetry *P* and time-reversal symmetry *T*, each band represents a Kramers doublet which makes the crossing point fourfold degenerate. These properties are the characteristic features of the type-II Dirac fermion and the crossing point is the well-known DP. Due to the $C_4$ four-fold rotation symmetry, there are four equivalent DPs in the 2D BZ [red dots in the inset of Fig. 2(a)]. The type-II DP due to the unavoidable crossing between the two $e_g$ bands was also found in the bulk cuprate oxides[18].

The DPs are protected by the mirror symmetry. Along Γ-M, wave vector **k** is invariant under the symmetry operation $M_{xy}$, which lies perpendicular to the $k_y = -k_x$ axis. In the spinless case, $M_{xy}^2 = 1$ and hence bands along Γ-M can be denoted by the eigenvalues (±1) of $M_{xy}$. It is evident that $M_{xy}\left|d_{x^2-y^2}\right\rangle = -\left|d_{x^2-y^2}\right\rangle$ and $M_{xy}\left|d_{z^2}\right\rangle = -\left|d_{z^2}\right\rangle$ (here $|\cdots\rangle$ denotes the orbital basis). The different eigenvalues (different irreducible representations) forbid the hybridization between the $d_{x^2-y^2}$ and $d_{z^2}$ bands. The DP is therefore protected by the $M_{xy}$ symmetry and cannot be gapped.

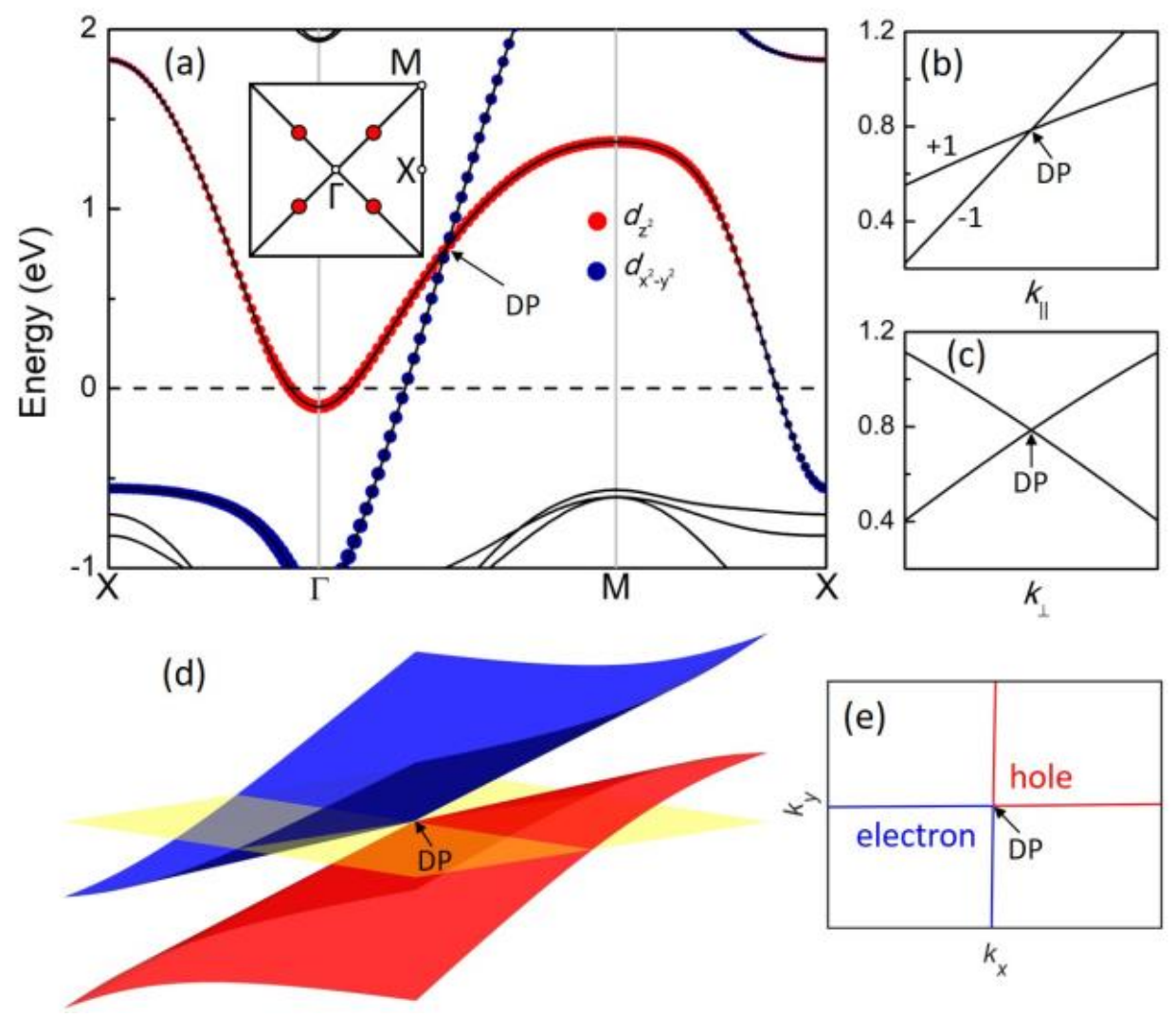

FIG. 2. (a) Band structure of the LAO/(LNO)$_1$/LAO along the high symmetry lines X(π, 0)–Γ(0, 0)–M(π, π)–X(π, 0) without SOC. Projection onto the $e_g$ ($d_{x^2-y^2}$, $d_{z^2}$) orbitals is indicated by dots which size corresponds to the weight of each orbital. Inset: the 2D BZ. Positions of DPs are indicated by the red dots. Zoom-in band structure around the DP (b) along Γ-M ($k_{\parallel}$) and (c) perpendicular to Γ-M ( $k_{\perp}$ ). (d) 3D band structure around the DP. (e) Energy contour for the isoenergy cutting [yellow plane in (d)] through the DP.

To gain further insight into the nature of the DP, we derive the $\mathbf{k}\cdot\mathbf{p}$ effective Hamiltonian based on the theory of invariants[36,37,38]. The Hamiltonian up to the linear order is given by[34]

$$H = \alpha_1\left(q_x + q_y\right)\sigma_0 + \beta_1\left(q_x - q_y\right)\sigma_x + \gamma_1\left(q_x + q_y\right)\sigma_z, \quad (1)$$

where $\left(q_x, q_y\right)$ are the wave vector deviation from DP and the $(\alpha_1, \beta_1, \gamma_1)$ expansion coefficients, $\sigma_0$ the 2×2 unitary matrix, and $(\sigma_x, \sigma_z)$ the Pauli matrices acting within the orbital spaces. The energy spectrum of Eq. (1) is

$$\varepsilon_{\pm} = \alpha_1\left(q_x + q_y\right) \pm \sqrt{\beta_1^2\left(q_x - q_y\right)^2 + \gamma_1^2\left(q_x + q_y\right)^2}. \quad (2)$$

Eq. (2) describes the 2D anisotropic massless Dirac fermions. Fitting to the band around the DP yields $\alpha_1 = 2.33$ eV Å, $\beta_1 = 2.13$ eV Å, and $\gamma_1 = 1.03$ eV Å. Along Γ-M, $q_x = q_y$ and the Dirac cone is strongly titled due to $\alpha_1 > \gamma_1$.

We see that the LAO/(LNO)$_1$/LAO hosts the 2D type-II Dirac fermions due to the symmetry protected band crossing between the two $e_g$ bands. However, the energy position of the DP is relatively high, i.e. ~0.8 eV above $E_F$. Electron doping or gate biasing are technically feasible to tune the DP closer to $E_F$. However, there is another efficient way to tune the DP by changing the quantum-well width[34] (LNO layer thickness).

Fig. 3(a) shows the calculated band structure of the LAO/(LNO)$_2$/LAO (001). Comparison to Fig. 2(a) reveals that the quantum confinement in the $z$ direction splits the two $d_{z^2}$ bands: at the Γ point, one band shifts above $E_F$ while the other band moves below $E_F$. On the other hand, the dispersion of the $d_{x^2-y^2}$ bands is nearly unaffected ($d_{x^2-y^2}$ orbitals lying in the $xy$-plane). As a result, the two crossing points between the lower $d_{z^2}$ band and the two $d_{x^2-y^2}$ bands become closer to $E_F$. These crossing points, denoted by P and Q, are seen in Fig. 3(b) along Γ-M ($k_{\|}$). Figs. 3(c) and 3(d) show the band dispersions perpendicular to Γ-M ($k_{\perp}$) and passing through the P and Q points, respectively. As can be seen, P point exhibits properties similar to the DP in the LAO/(LNO)$_1$/LAO: the linear dispersion in all directions, the titled band crossing along $k_{\|}$ [Fig. 3(b)], and the straight band crossing along $k_{\perp}$ [Fig. 3(c)]. Q point reveals however a different behavior: along $k_{\|}$, the crossing bands have the titled linear dispersion [Fig. 3(b)], while along $k_{\perp}$, the dispersion is quadratic [Fig. 3 (d)]. Using a tight-binding (TB) model (discussed below), we have calculated the band dispersion around $E_F$. Fig. 4(a) shows that P point emerges as an isolated point, while Q point belongs to a closed nodal line (NL). It is evident from Fig. 4(b) that the energy of the NL varies from –0.04 eV to –0.7 eV, which favors the experimental measurement over a large energy scale.

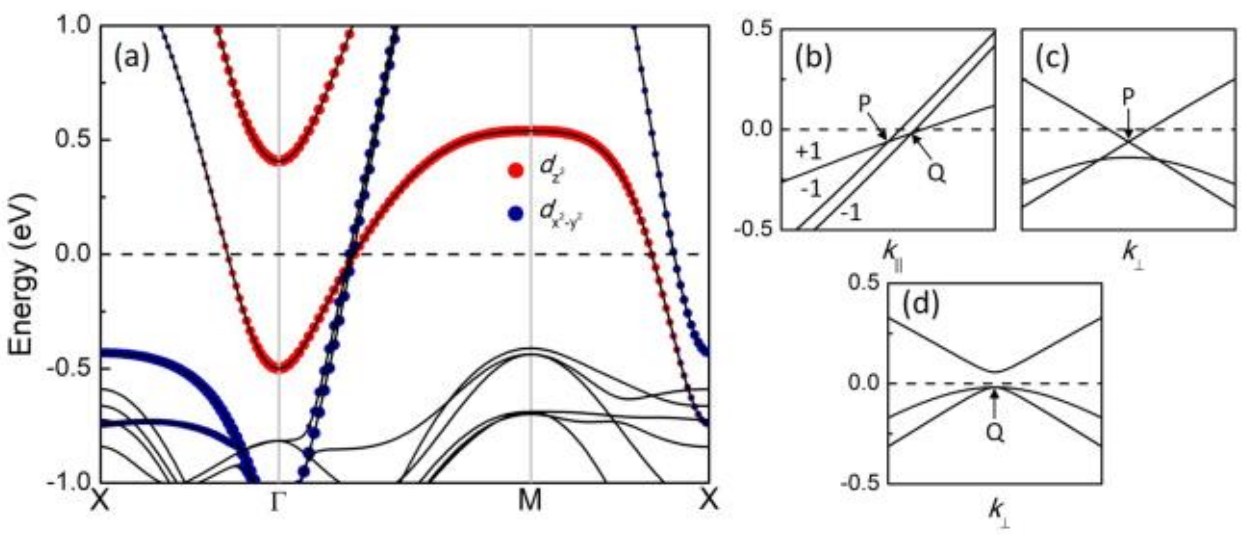

FIG. 3. (a) Band structure of the LAO/(LNO)$_2$/LAO without SOC. Projection onto the $e_g$ ($d_{x^2-y^2}$, $d_{z^2}$) orbitals is indicated by dots which size corresponds to the weight of each orbital. Zoom-in band structure around crossing points (b) along Γ-M ($k_{\|}$) and (c, d) perpendicular to Γ-M ($k_{\perp}$) passing through (c) P and (d) Q points.

The NL appears due to the $PT$ symmetry, which in general produces a closed nodal loop in the BZ[39,40,41]. The point group for a general crossing point at the NL is $C_s$, which consists of the identity element $E$ and the mirror reflection $M_z$ with respect to the plane perpendicular to the $z$ axis. There are two irreducible representations of the $C_s$ group[42], A′ and A″. Examining the two bands forming the NL [Fig. S5a], we find that one band belongs to the A′ representation, while the other one belongs to the A″ representation[34]. Since the eigenvalue of $M_z$ for the A′(A″) representation is +1 (–1)[42,43], the NL is protected by the mirror symmetry $M_z$.

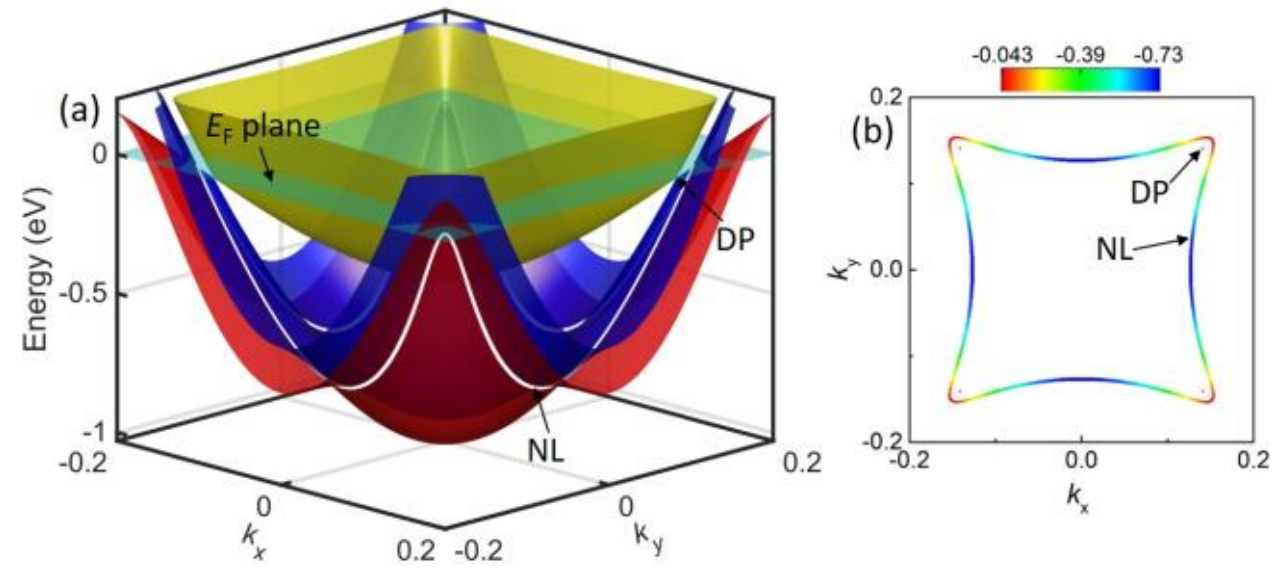

FIG. 4. (a) 3D plot of the band structure around the Fermi energy. Band touching points between red and blue branches form the closed node line (NL), as indicated by white loop. (b) 2D projection of the isolated DPs and the closed NL. $k_x$ and $k_y$ are in units of $2\pi/a$. The color map quantifies the band energies at the touching points. The results are obtained using the tight-binding Hamiltonian model as described in text.

To obtain a deeper insight into the band structure of the LAO/LNO/LAO, we construct a TB Hamiltonian in basis of the $\left|d_{x^2-y^2}\right\rangle$ and $\left|d_{z^2}\right\rangle$ orbitals, which are denoted as $\left|\alpha\right\rangle$ and $\left|\beta\right\rangle$, respectively. The Hamiltonian model and energy spectrum can be found in the Supplementary Material[34]. Along Γ-M ($k_x = k_y$), the band energies are $\varepsilon_{\pm} = h_\beta, h_\alpha$ for one-u.c.-thick LNO and $\varepsilon_{1,4} = h_\beta \mp h_{\beta z}$, $\varepsilon_{2,3} = h_\alpha \mp h_{\alpha z}$ for two-u.c.-thick LNO [supplementary Eqs. (5) and (12)], where $h$'s are the functions of $\left(k_x, k_y\right)$. In the latter case, both $h_\alpha$ and $h_\beta$ are split into two subbands with the splitting energy being proportional to $2h_{\alpha z}$ and $2h_{\beta z}$, respectively. Since the $\left|\alpha\right\rangle$ ($\left|\beta\right\rangle$) orbital lies in (out of) the plane, the out-of-plane hopping $h_{\alpha z}$ is expected to be much smaller than $h_{\beta z}$, which is confirmed by the fitting[34]. Consequently, the splitting between the two $d_{z^2}$ bands is significantly larger than the splitting between the two $d_{x^2-y^2}$ bands, which is in line with our DFT results [Fig. 3(a)].

It is noteworthy that the previous experimental work[44,45] has demonstrated a metal-insulator transition (MIT) in a few-u.c.-thick LNO film due to the low dimensionality and strain. However, very recently it was demonstrated that oxygen vacancies play a critical role in triggering a MIT[46]. Moreover, both the DFT and DFT+DMFT (dynamical mean field theory) results reveal the metallic feature even for one-u.c.-thick LNO film without oxygen vacancies[46], in agreement with our results.

Finally, we discuss the effect of SOC on the DPs from symmetry arguments. Here we consider a LAO/$(LNO)_1$/LAO as an example, but the same conclusion applies to LAO/$(LNO)_2$/LAO as well. Along Γ-M, there are two symmetry invariant operations, i.e. $PT$ and $M_{xy}$. In real space, it can be easily checked that $\left[PT, M_{xy}\right]=0$. In spin space, we have $PT = i\sigma_y K$ and $M_{xy} = i\left(\sigma_x - \sigma_y\right)/\sqrt{2}$. Combining real and spin space, we obtain $\left[PT, M_{xy}\right]=0$ and $M_{xy}^2 = -1$. Along Γ-M, the Bloch states $\left|\psi\right\rangle$ and $PT\left|\psi\right\rangle$ can be labeled using the eigenvalues of $M_{xy}$, namely $M_{xy}\left|\psi\right\rangle = \pm i\left|\psi\right\rangle$ and $M_{xy}PT\left|\psi\right\rangle = \mp iPT\left|\psi\right\rangle$. The doubly degenerate states $\left|\psi\right\rangle$ and $PT\left|\psi\right\rangle$ have opposite $M_{xy}$ eigenvalues. Therefore, when two sets of such doublet bands cross at the DP, the two bands with the same $M_{xy}$ eigenvalue can hybridize and open a gap at the DP, as schematically shown in Figs. S3(b) and S3(c). To confirm this, we have calculated the band structure by including SOC. As shown in the Fig. S4, a tiny gap of ~2 meV appears at the DP. Thus, in the spinful case, the DP is not protected by mirror symmetry and a small gap opens due to SOC. The same conclusion also applies to the NL [Fig. S5(b)].

***Summary.*** In summary**,** we have predicted that LAO/LNO/LAO (001) hosts the 2D type-II Dirac fermions. Our DFT calculations and TB modeling demonstrate that one-u.c.-thick LNO layer, when placed in the quantum well, forms the Dirac points that are located along the diagonal lines of the 2D Brillouin zone and in the spinless case are protected by the mirror symmetry. The formation of the 2D type-II Dirac fermions can be tuned by the quantum-well width. Moreover, growing the LAO/LNO/LAO on different substrates [47, 48] enables the strain-tunable DP[34]. For a two-u.c.-thick LNO layer, the energy position of the Dirac point is found to appear near the Fermi energy. In this case, we predict the coexistence of the isolated Dirac points and a closed nodal line. SOC opens a small (~2 meV) gap at the Dirac points and the nodal line. We hope that our findings will stimulate the experimental search for the 2D type-II Dirac fermions in the LAO/LNO/LAO (001) quantum-well structures.

***Acknowledgments.*** This work was supported by the National Science Foundation (NSF) through Nebraska Materials Research Science and Engineering Center (MRSEC) (NSF Grant No. DMR-1420645). Computations were performed at the University of Nebraska Holland Computing Center. The atomic structure was produced using VESTA software[49].

[*] E-mail: ltao2@unl.edu

[†] E-mail: tsymbal@unl.edu

# SUPPLEMENTAL MATERIAL

## Two-dimensional type-II Dirac fermions in a $LaAlO_3/LaNiO_3/LaAlO_3$ quantum well

L. L. Tao, and Evgeny Y. Tsymbal

*Department of Physics and Astronomy & Nebraska Center for Materials and Nanoscience*
*University of Nebraska, Lincoln, Nebraska 68588, USA*

## 1. Computational model and details

An energy cutoff of 680 eV and local density approximation (LDA)[1] for the exchange and correlation functional are used throughout. The LAO/LNO/LAO quantum-well structure is modelled using $(LNO)_n/(LAO)_m$ superlattices, where $n$ and $m$ refer to the number of unit cells (u.c.) of LNO and LAO, respectively. Calculations are performed for $n$ = 1-4 and $m$ = 6. The latter was found to be sufficient to eliminate the interaction between the quantum wells in the superlattice. The in-plane lattice constant $a$ of the superlattices is fixed to the calculated lattice constant of bulk cubic LAO, $a$=3.757 Å, to simulate epitaxial growth on a LAO substrate. The out-of-plane lattice constant as well as the atomic positions are fully relaxed with the force tolerance of 2.6 meV/Å. Previous work[2] shows that the conventional LDA method gives the best agreement with the experimental data for LNO. As a comparison, the Hubbard $U$ correction for Ni-$d$ electron is included[3] to consider the correlation effects, as presented in section 4.

## 2. Tight-binding model

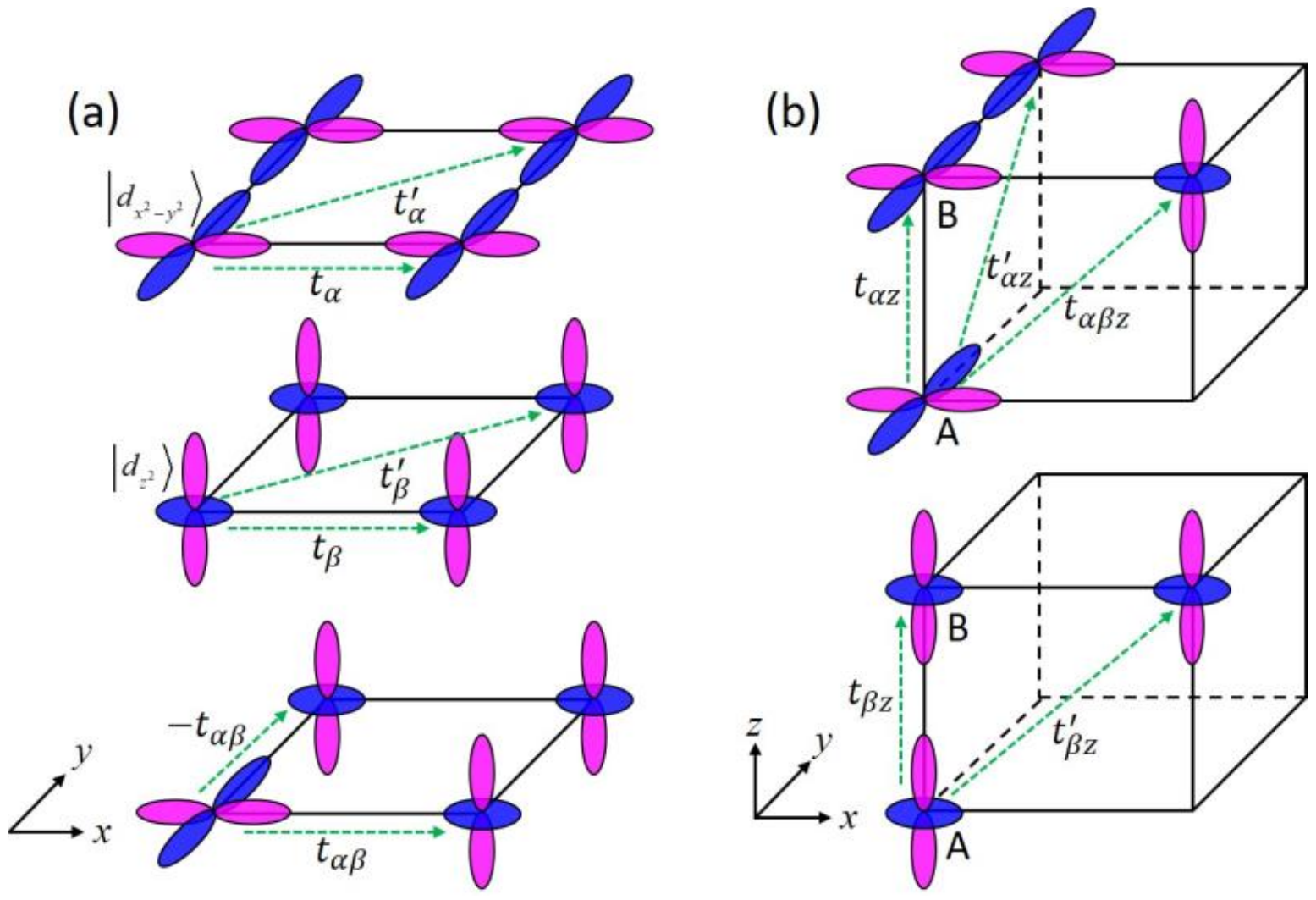

**FIG. S1.** Sketch of hopping (green dashed lines) between different orbitals. There are two orbitals on each site (a) for $(LNO)_1/(LAO)_6$ and (b) for $(LNO)_2/(LAO)_6$. In (b), A and B denote two Ni sites.

The TB Hamiltonian for $(LNO)_1/(LAO)_6$ in real space reads

$$\mathcal{H}_1 = \sum_{n_x,n_y}\left(\mathbf{C}^\dagger_{n_x,n_y}V_0\mathbf{C}_{n_x,n_y} + \mathbf{C}^\dagger_{n_x,n_y}T_x\mathbf{C}_{n_x+1,n_y} + \mathbf{C}^\dagger_{n_x,n_y}T_y\mathbf{C}_{n_x,n_y+1} + \mathbf{C}^\dagger_{n_x,n_y}T_{xy}\mathbf{C}_{n_x+1,n_y+1} + \mathbf{C}^\dagger_{n_x,n_y}T_{xy}\mathbf{C}_{n_x-1,n_y+1} + \text{H.c.}\right), \qquad (1)$$

where ( $n_x, n_y$ ) is the lattice site index, $\mathbf{C}_{n_x,n_y} = \left(c_{\alpha n_x,n_y}, c_{\beta n_x,n_y}\right)^{\mathrm{T}}$ (T denotes transpose) is the second-quantized annihilation operator, and H.c. refers to the Hermitian conjugate. Here we use $\alpha$ ( $\beta$ ) to denote the $\left|d_{x^2-y^2}\right\rangle$ ( $\left|d_{z^2}\right\rangle$ ) orbital for convenience of notation. The onsite ($V_0$) and hopping ($T$'s) matrices can be expressed as

$$V_0 = \begin{bmatrix} \varepsilon_\alpha & 0 \\ 0 & \varepsilon_\beta \end{bmatrix},\ T_x = \begin{bmatrix} t_\alpha & t_{\alpha\beta} \\ t_{\alpha\beta} & t_\beta \end{bmatrix},\ T_y = \begin{bmatrix} t_\alpha & -t_{\alpha\beta} \\ -t_{\alpha\beta} & t_\beta \end{bmatrix},\ T_{xy} = \begin{bmatrix} t'_\alpha & 0 \\ 0 & t'_\beta \end{bmatrix}, \tag{2}$$

here $\varepsilon_\alpha$ ( $\varepsilon_\beta$ ) denotes the onsite energy, $t_\alpha$ ( $t_\beta$ ) the nearest-neighbor (NN) intra-orbital hopping, $t'_\alpha$ ( $t'_\beta$ ) the next-nearest-neighbor (NNN) intra-orbital hopping, and $t_{\alpha\beta}$ the NN inter-orbital hopping. Those hopping parameters are schematically shown in Fig. S1(a). By performing Fourier transformations $\mathbf{C}_{n_x,n_y} = \frac{1}{\sqrt{N}}\sum_k \mathbf{C}_k e^{ik_x n_x + ik_y n_y}$ , where $\mathbf{C}_k = \left(c_{\alpha k}, c_{\beta k}\right)^{\mathrm{T}}$ and the lattice constants have been assumed to be units for convenience of notation, we then obtain $\mathcal{H}_1 = \sum_k \mathbf{C}_k^\dagger H_1 \mathbf{C}_k$ and $H_1$ reads

$$H_1 = \frac{1}{2}\left(h_\alpha + h_\beta\right)\sigma_0 + \frac{1}{2}\left(h_\alpha - h_\beta\right)\sigma_z + h_{\alpha\beta}\sigma_x, \tag{3}$$

where $\sigma_0$ is the 2×2 unitary matrix, and $(\sigma_x, \sigma_z)$ are the Pauli matrices acting within the ( $\left|d_{x^2-y^2}\right\rangle, \left|d_{z^2}\right\rangle$ ) orbital space. The elements $h$'s are defined as

$$\begin{cases} h_\alpha = \varepsilon_\alpha + 2t_\alpha\left(\cos k_x + \cos k_y\right) + 4t'_\alpha \cos k_x \cos k_y \\ h_\beta = \varepsilon_\beta + 2t_\beta\left(\cos k_x + \cos k_y\right) + 4t'_\beta \cos k_x \cos k_y, \\ h_{\alpha\beta} = 2t_{\alpha\beta}\left(\cos k_x - \cos k_y\right) \end{cases} \tag{4}$$

The energy spectrum of Eq. (1) is

$$\varepsilon_\pm = \frac{1}{2}\left(h_\alpha + h_\beta\right) \pm \frac{1}{2}\sqrt{\left(h_\alpha - h_\beta\right)^2 + 4h_{\alpha\beta}^2}. \tag{5}$$

Analogously, the TB Hamiltonian for $(LNO)_2/(LAO)_6$ in real space reads

$$\mathcal{H}_2 = \tau_0 \otimes \mathcal{H}_1 + \sum_{n_x,n_y}\left(\mathbf{C}_{n_x,n_y}^\dagger T_z \mathbf{C}_{n_x,n_y} + \mathbf{C}_{n_x,n_y}^\dagger T_{xz} \mathbf{C}_{n_x+1,n_y} + \mathbf{C}_{n_x,n_y}^\dagger T_{yz} \mathbf{C}_{n_x,n_y+1} + \mathrm{H.c.}\right), \tag{6}$$

where $\mathbf{C}_{n_x,n_y} = \left(c_{\alpha n_x,n_y}^{\mathrm{A}}, c_{\beta n_x,n_y}^{\mathrm{A}}, c_{\alpha n_x,n_y}^{\mathrm{B}}, c_{\beta n_x,n_y}^{\mathrm{B}}\right)^{\mathrm{T}}$ , A and B denotes two Ni sites, as shown in Fig. S1(b). The inter-site hopping matrices can be expressed as

$$T_z = \frac{1}{2}\left(\tau_x + i\tau_y\right)\otimes\begin{bmatrix} t_{\alpha z} & 0 \\ 0 & t_{\beta z} \end{bmatrix}, T_{xz} = \frac{1}{2}\left(\tau_x + i\tau_y\right)\otimes\begin{bmatrix} t'_{\alpha z} & t_{\alpha\beta z} \\ t_{\alpha\beta z} & t'_{\beta z} \end{bmatrix}, T_{yz} = \frac{1}{2}\left(\tau_x + i\tau_y\right)\otimes\begin{bmatrix} t'_{\alpha z} & -t_{\alpha\beta z} \\ -t_{\alpha\beta z} & t'_{\beta z} \end{bmatrix}, \tag{7}$$

where $t_{\alpha z}$ ( $t_{\beta z}$ ) denotes the NN intra-orbital inter-site hopping, $t'_{\alpha z}$ ( $t'_{\beta z}$ ) the NNN intra-orbital inter-site hopping, and $t_{\alpha\beta z}$ the NNN inter-orbital inter-site hopping. To describe the degree of freedom for two Ni sites, we introduce another set of Pauli matrices τ's. By performing Fourier transformations, we get $\mathcal{H}_2 = \sum_k \mathbf{C}_k^\dagger H_2 \mathbf{C}_k$ , where $\mathbf{C}_k = \left(c_{\alpha k}^{\mathrm{A}}, c_{\beta k}^{\mathrm{A}}, c_{\alpha k}^{B}, c_{\beta k}^{\mathrm{B}}\right)^{\mathrm{T}}$ and $H_2$ reads

$$H_2 = \tau_0 \otimes H_1 + \frac{1}{2}\tau_x \otimes \left[ \left( h_{\alpha z} + h_{\beta z} \right)\sigma_0 + \left( h_{\alpha z} - h_{\beta z} \right)\sigma_z + 2h_{\alpha\beta z}\sigma_x \right], \tag{8}$$

where the additional elements are defined as

$$\begin{cases} h_{\alpha z} = t_{\alpha z} + 2t'_{\alpha z}\left(\cos k_x + \cos k_y\right) \\ h_{\beta z} = t_{\beta z} + 2t'_{\beta z}\left(\cos k_x + \cos k_y\right), \\ h_{\alpha\beta z} = 2t_{\alpha\beta z}\left(\cos k_x - \cos k_y\right) \end{cases} \tag{9}$$

Since the Hamiltonian preserves the $P$ (inversion) symmetry, which is given by $P = \tau_x \otimes \sigma_0$ for $(LNO)_2/(LAO)_6$ in spinless case, we have $[\tau_x \otimes \sigma_0, H_2] = 0$, using the unitary matrix $U$ which diagonalizes $\tau_x \otimes \sigma_0$, $H_2$ can be transformed into a block diagonal. $U$ is given by

$$U = \frac{1}{\sqrt{2}}\begin{bmatrix} -1 & 0 & 0 & -1 \\ 0 & 1 & 1 & 0 \\ 1 & 0 & 0 & -1 \\ 0 & -1 & 1 & 0 \end{bmatrix}, \text{ and } U^\dagger \tau_x \sigma_0 U = \begin{bmatrix} -1 & 0 & 0 & 0 \\ 0 & -1 & 0 & 0 \\ 0 & 0 & 1 & 0 \\ 0 & 0 & 0 & 1 \end{bmatrix}. \tag{10}$$

Then we have

$$\tilde{H}_2 = U^\dagger H_2 U = \begin{bmatrix} h_\alpha - h_{\alpha z} & h_{\alpha\beta z} - h_{\alpha\beta} & 0 & 0 \\ h_{\alpha\beta z} - h_{\alpha\beta} & h_\beta - h_{\beta z} & 0 & 0 \\ 0 & 0 & h_\beta + h_{\beta z} & -h_{\alpha\beta z} - h_{\alpha\beta} \\ 0 & 0 & -h_{\alpha\beta z} - h_{\alpha\beta} & h_\alpha + h_{\alpha z} \end{bmatrix}. \tag{11}$$

The energy spectrum of Eq. (11) is

$$\begin{cases} \varepsilon_{1,2} = \frac{1}{2}\left(h_\alpha + h_\beta - h_{\alpha z} - h_{\beta z}\right) \pm \frac{1}{2}\sqrt{\left(h_\alpha - h_\beta - h_{\alpha z} + h_{\beta z}\right)^2 + 4\left(h_{\alpha\beta} - h_{\alpha\beta z}\right)^2} \\ \varepsilon_{3,4} = \frac{1}{2}\left(h_\alpha + h_\beta + h_{\alpha z} + h_{\beta z}\right) \pm \frac{1}{2}\sqrt{\left(h_\alpha - h_\beta + h_{\alpha z} - h_{\beta z}\right)^2 + 4\left(h_{\alpha\beta} + h_{\alpha\beta z}\right)^2} \end{cases}. \tag{12}$$

The fitting parameters are summarized in Table S1 and the comparison between DFT and TB results are shown in Fig. S2, from which the overall agreement can be seen.

**Table S1.** List of fitting TB parameters (unit: eV) for both $(LNO)_1/(LAO)_6$ and $(LNO)_2/(LAO)_6$.

| $(LNO)_1/(LAO)_6$ | | $(LNO)_2/(LAO)_6$ | | |
|---|---|---|---|---|
| $\varepsilon_\alpha = 1.1993$ | $\varepsilon_\beta = 0.9407$ | $\varepsilon_\alpha = 1.2147$ | $\varepsilon_\beta = 0.9413$ | $t_\alpha = -0.4918$ |
| $t_\alpha = -0.4922$ | $t_\beta = -0.1741$ | $t_\beta = -0.1614$ | $t'_\alpha = -0.0584$ | $t'_\beta = -0.0695$ |
| $t'_\alpha = -0.0624$ | $t'_\beta = -0.0720$ | $t_{\alpha z} = 0.0269$ | $t_{\beta z} = -0.6182$ | $t'_{\alpha z} = 0.0043$ |
| $t_{\alpha\beta} = 0.4823$ | | $t'_{\beta z} = 0.0441$ | $t_{\alpha\beta} = 0.4826$ | $t_{\alpha\beta z} = -0.0397$ |

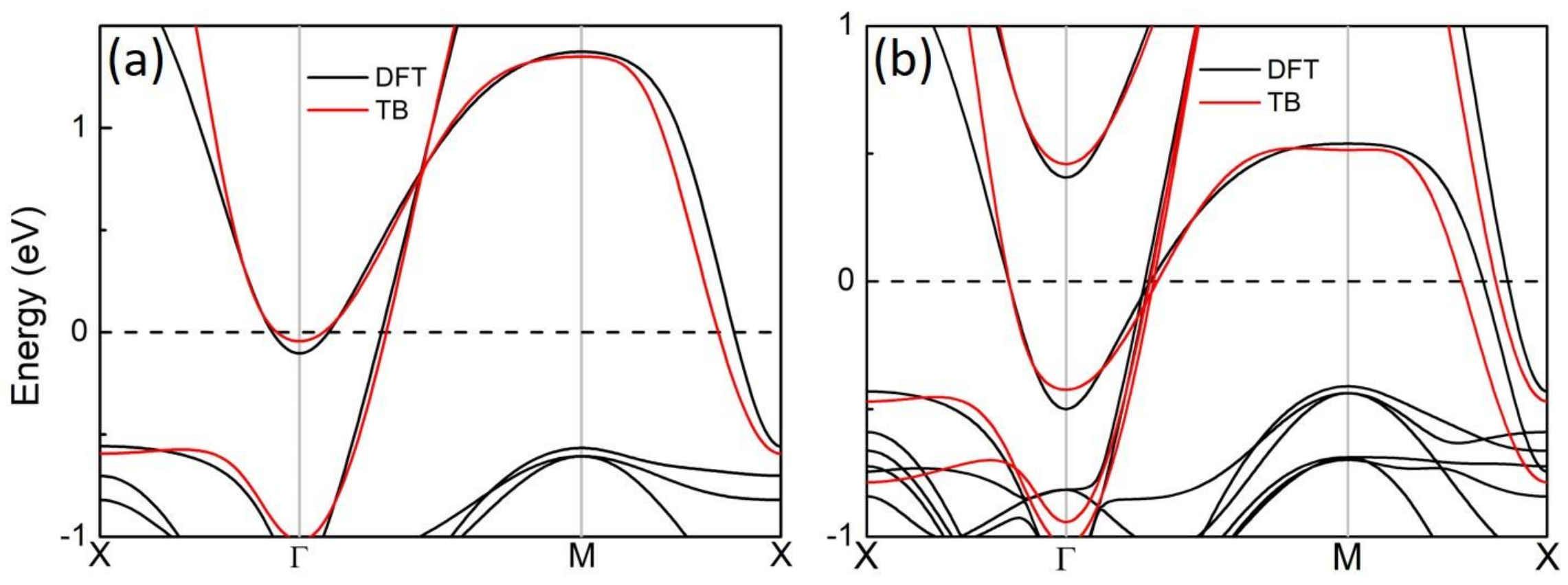

**FIG. S2.** Comparison between the DFT and TB band structures for (a) $(LNO)_1/(LAO)_6$ and (b) $(LNO)_2/(LAO)_6$.

## 3. Effect of SOC

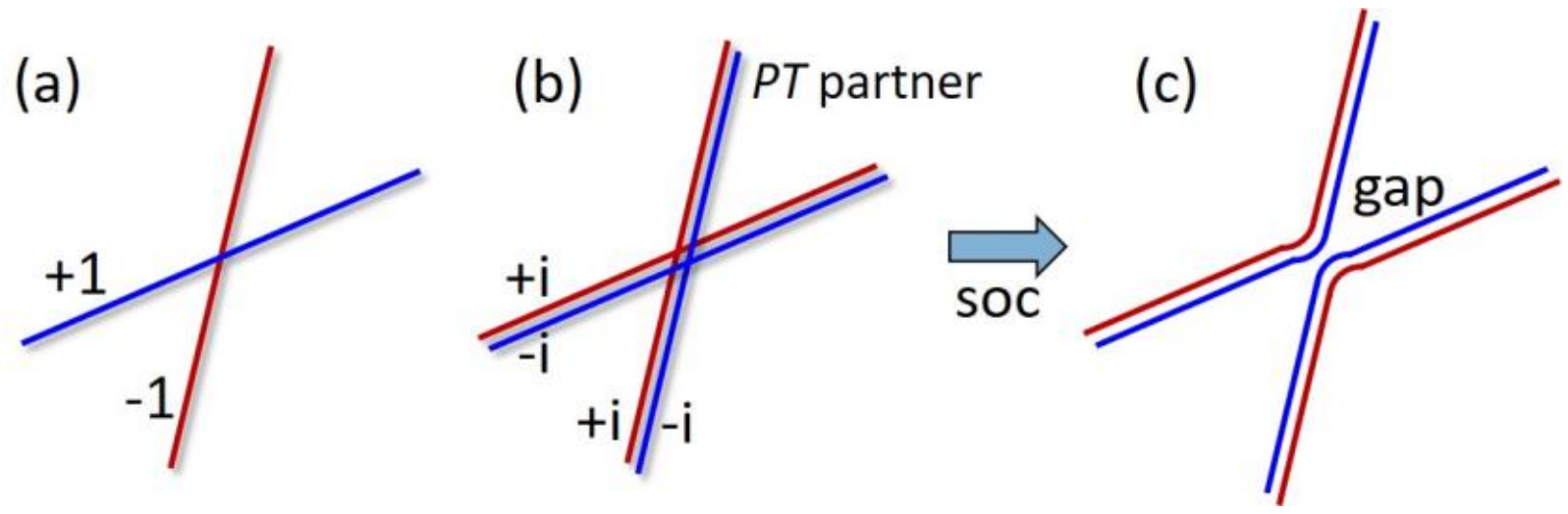

**FIG. S3.** Schematic band dispersion along the Γ-M line around the DP for (a) spinless case and (b, c) spinfull case. The red and blue colors denote opposite eigenvalues of symmetry $M_{xy}$ . In (b) and (c), the double degenerate bands are slightly offset to indicate their opposite eigenvalues.

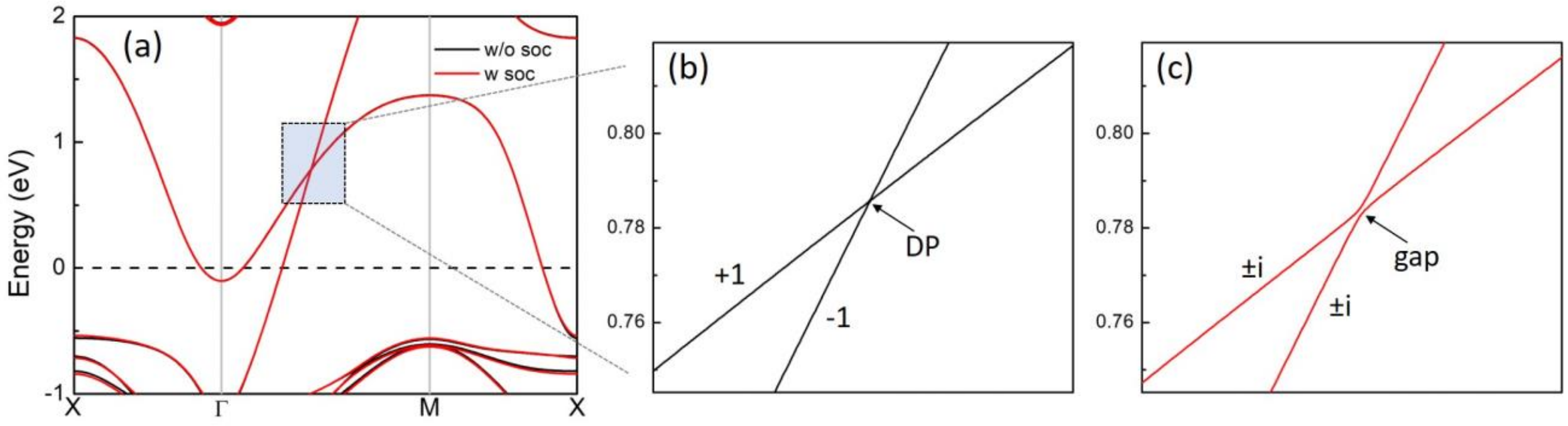

**FIG. S4.** (a) Band structures for $(LNO)_1/(LAO)_6$ with (w) SOC and without (w/o) SOC. (b, c) Zoom-in around the DP, as highlighted by the rectangular box in (a). (b) without SOC and (c) with SOC. The eigenvalues of mirror $M_{xy}$ are labeled on each band.

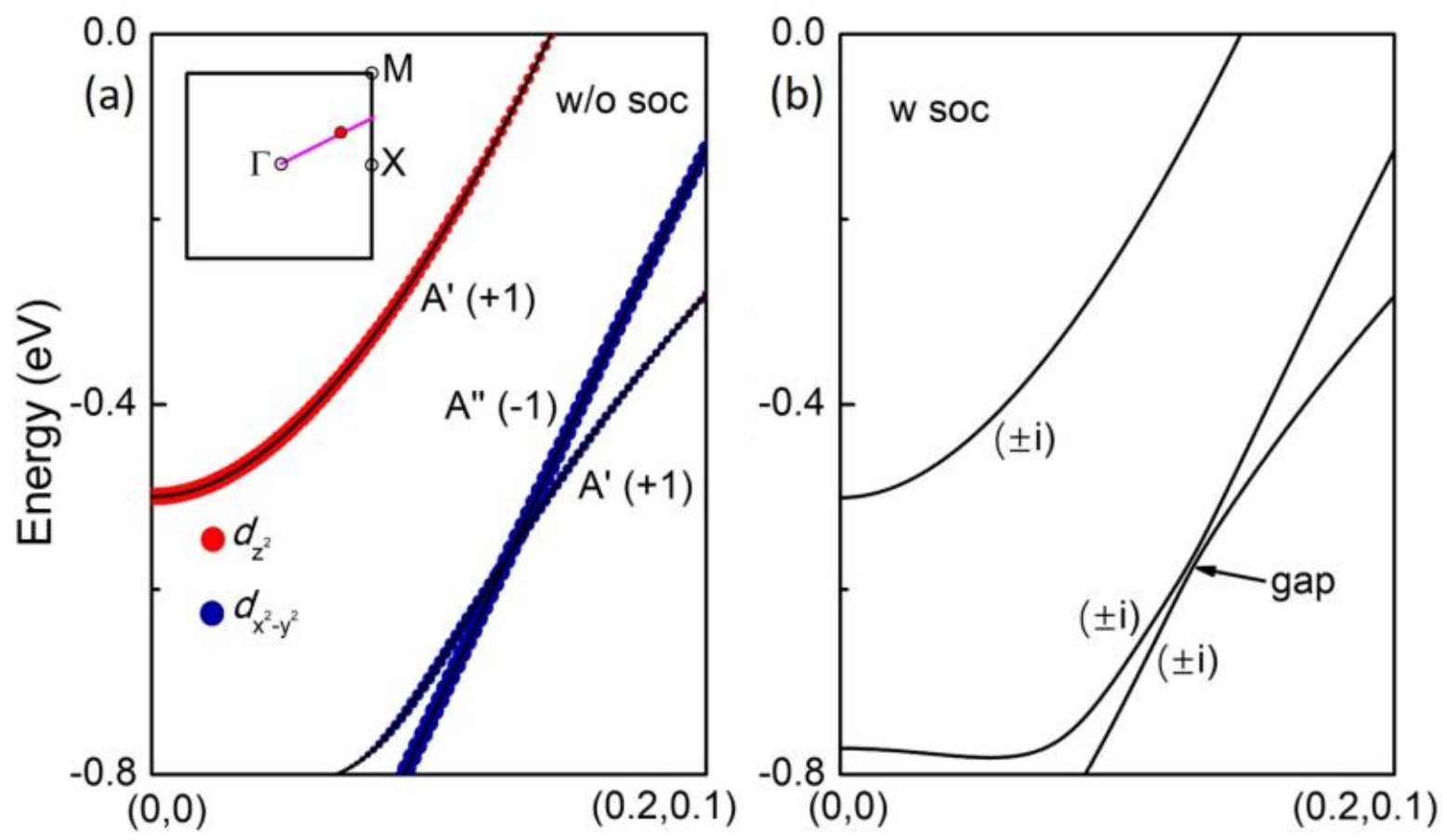

**FIG. S5.** Band structures along general *k* path (inset) crossing the node line. The crossing point is indicated by the solid red dot. (a) without (w/o) SOC and (b) with (w) SOC. The irreducible representations for each band are denoted and the number in parentheses represents the eigenvalue (character) of mirror $M_z$.

Fig. S5 shows the band structure along general *k* path. With SOC (Fig. S5b), the node line is gapped. In spinfull case (with SOC), we have $PT = i\sigma_y K$ and $M_z = i\sigma_z$. For the double degenerate states $|\psi\rangle$ and $PT|\psi\rangle$, we have $M_z|\psi\rangle = \pm i|\psi\rangle$ and $M_z PT|\psi\rangle = \mp iPT|\psi\rangle$. As seen from Fig. S3, the node line is therefore gapped by SOC and cannot be protected by $M_z$ symmetry in the spinfull case.

## 4. Band structure with LDA+U

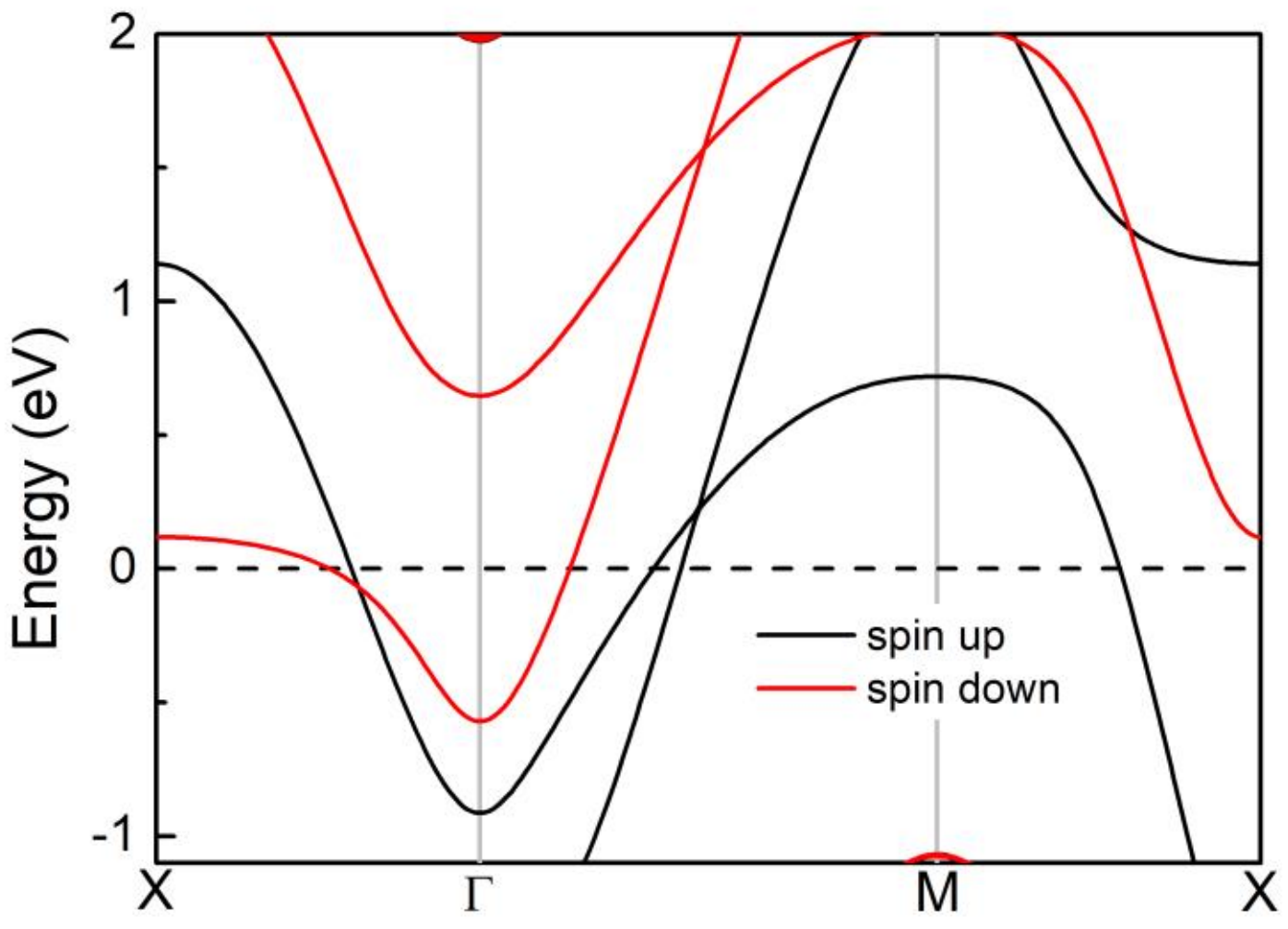

**FIG. S6.** Band structure of $(LNO)_1/(LAO)_6$ with $U$=5.5 eV in the absence of SOC.

The ferromagnetic state is found with $U$ = 5.5 eV or larger, which is similar to that (5.7 eV[4]) used previously. Fig. S6 shows the band structure for $(LNO)_1/(LAO)_6$ with $U$ = 5.5 eV, which yields the ferromagnetic state. The spin splitting is clearly seen. The DP due to the crossing between $d_{x^2-y^2}$ and $d_{z^2}$ bands still exists along Γ-M.

## 5. Effect of epitaxial strain

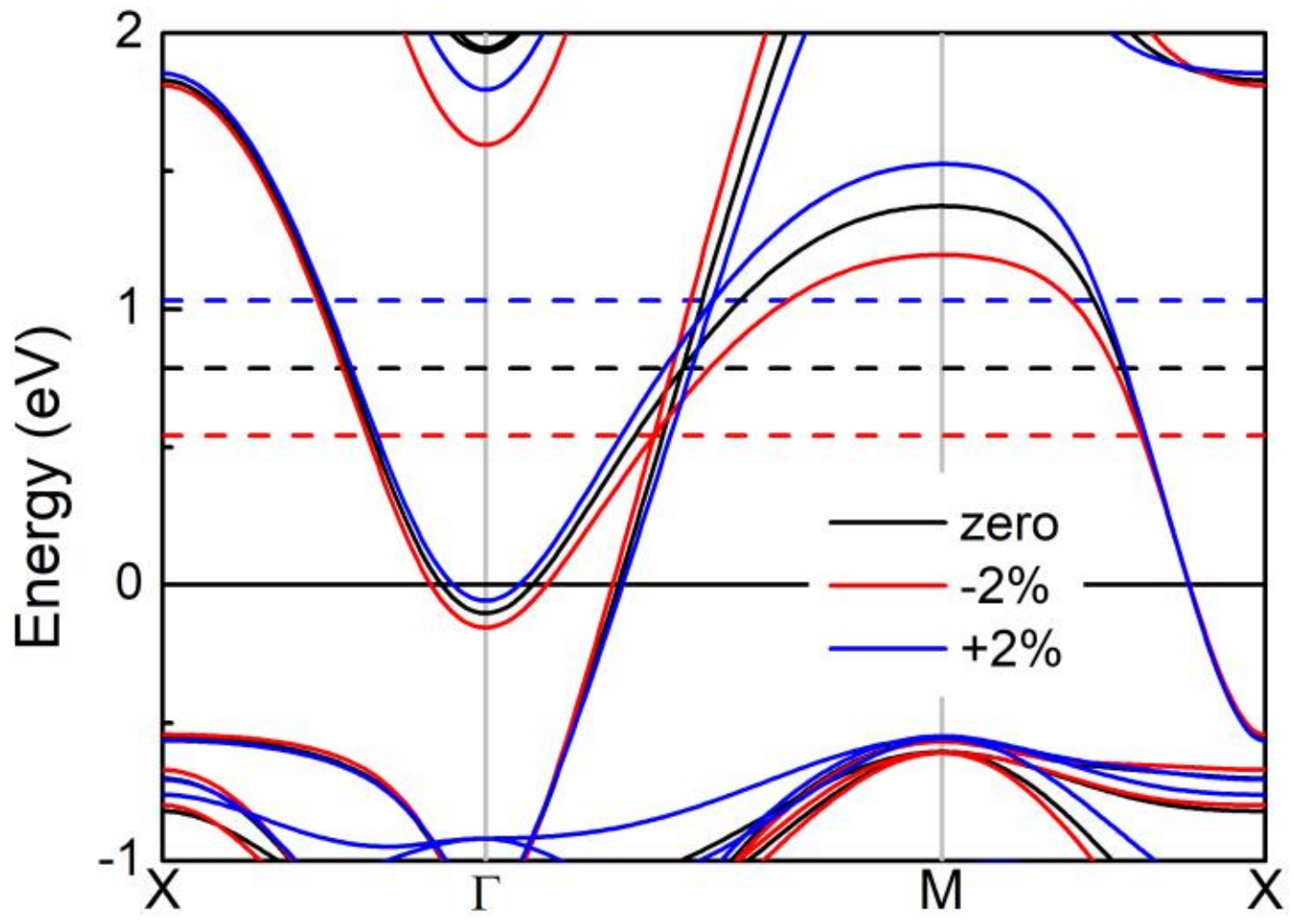

**FIG. S7.** Band structure of $(LNO)_1/(LAO)_6$ under different epitaxial strains without SOC. The dashed lines denote the energy positions of each DP.

The effects of epitaxial strain are studied through changing the in-plane lattice constant by ±2% while fully relaxing the out-of-plane lattice constant as well as the atomic coordinators. Since the experimental lattice constant for pseudo cubic LNO is *a*=3.84 Å[5], the compressive (–2%) and tensile (+2%) strain can be achieved through growing LNO/LAO on the tetragonal $LaSrAlO_4$ (*a*=3.75 Å[6]) and cubic $SrTiO_3$ (*a*=3.91 Å[6]) substrates, respectively. Fig. S7 shows the comparison of the band structures of $(LNO)_1/(LAO)_6$ under compressive (–2%) and tensile (+2%) strains. We see that the compressive (tensile) stain pushes the DP downward (upward) by ~0.24 eV, which enables the achievement of strain-tunable DP.

## 6. Effect of quantum-well width

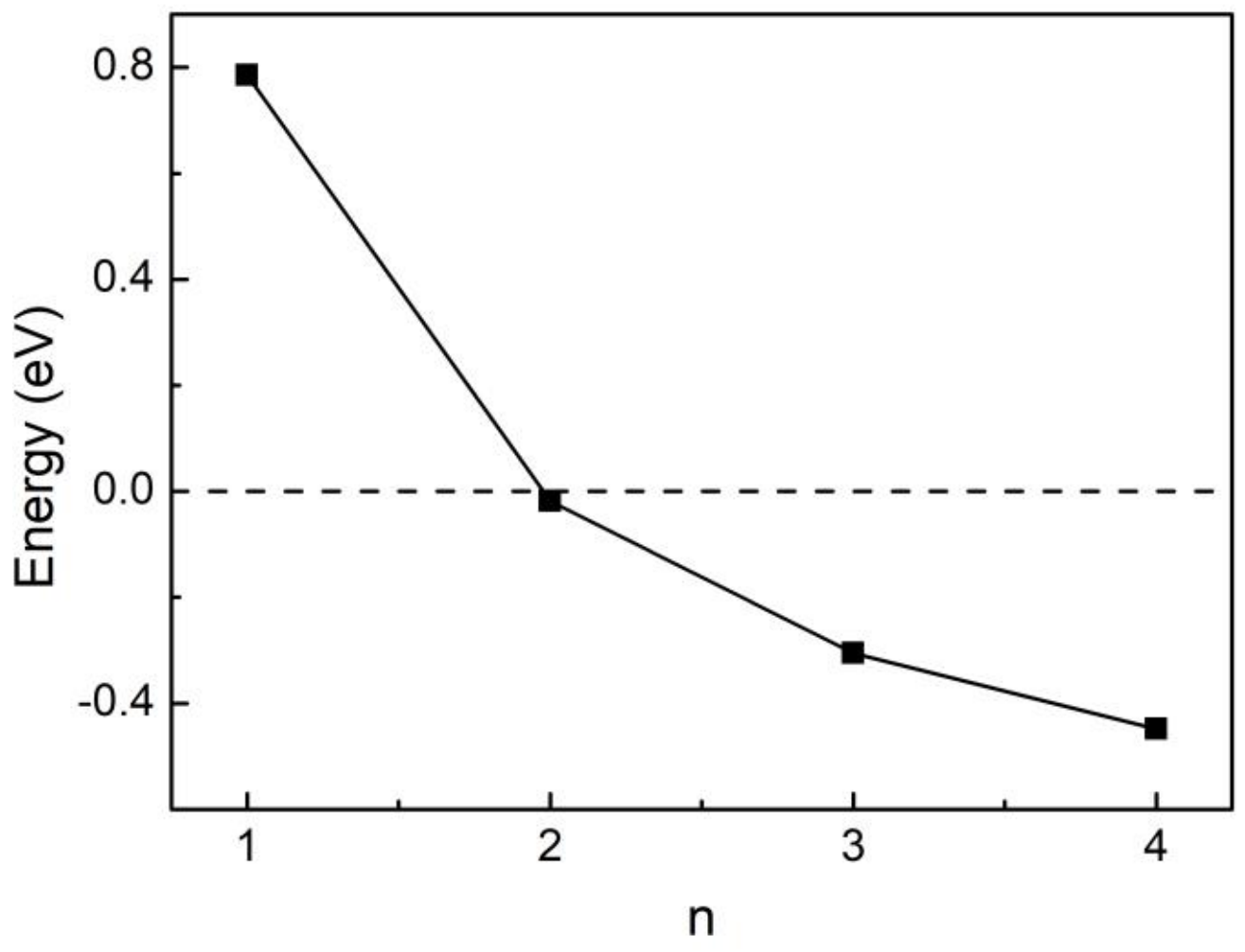

**FIG. S8.** The position of the Dirac point as a function of quantum-well width *n*, which denotes the number of unit cells of LNO in the LAO/(LNO)$_n$/LAO structure. For $n > 1$, we plotted the Dirac point that is nearest to the Fermi energy which lies at zero and is indicated by the dashed line.

Fig. S8 shows the calculated energy position of the Dirac points as a function of quantum-well width. It is seen that the position of the Dirac point is very close to the Fermi energy for a two-u.c.-thick LNO layer ($n$=2) and can be tuned by changing the quantum-well width.

## 7. $\mathbf{k}\cdot\mathbf{p}$ Hamiltonian

A $\mathbf{k}\cdot\mathbf{p}$ Hamiltonian can be derived based on theory of invariants[7,8,9]. The Hamiltonian around the DP $\mathbf{k_0}$ can be expanded as follows:

$$H(\mathbf{k}_0+\mathbf{q}) = E(\mathbf{k}_0) + d_0(\mathbf{q})\sigma_0 + \mathbf{d}(\mathbf{q})\cdot\boldsymbol{\sigma}, \tag{13}$$

where $E(\mathbf{k}_0)$ is a constant energy at the DP, $\sigma_0$ is the 2×2 unitary matrix, $\boldsymbol{\sigma}=(\sigma_x,\sigma_y,\sigma_z)$ represent the Pauli matrices acting within the ($|d_{x^2-y^2}\rangle$, $|d_{z^2}\rangle$) orbital space, $\mathbf{q}=(q_x,q_y,q_z)$ is the wave vector deviation from $\mathbf{k_0}$, and $\mathbf{d}(\mathbf{q})=(d_x(\mathbf{q}),d_y(\mathbf{q}),d_z(\mathbf{q}))$ is a vector function of $\mathbf{q}$. The symmetry operations at the DP are $PT$ and $M_{xy}$, which can be expressed as $PT=\sigma_0 K$ ($K$ represents the complex conjugation) and $M_{xy}=-\sigma_z$ within the ($|d_{x^2-y^2}\rangle$, $|d_{z^2}\rangle$) orbital basis. The functions $d_0(\mathbf{q})$ and $\mathbf{d}(\mathbf{q})$ due to the symmetry constraints satisfy the following conditions

$$\begin{aligned} PT &: \begin{cases} d_0(q_x,q_y) = d_0(q_x,q_y), \\ d_{x,z}(q_x,q_y) = d_{x,z}(q_x,q_y), \\ d_y(q_x,q_y) = -d_y(q_x,q_y), \end{cases} \\ M_{xy} &: \begin{cases} d_0(q_x,q_y) = d_0(q_y,q_x), \\ d_{x,y}(q_x,q_y) = -d_{x,y}(q_y,q_x), \\ d_z(q_x,q_y) = d_z(q_y,q_x). \end{cases} \end{aligned} \tag{14}$$

Based on these conditions, the symmetry allowed terms up to the cubic order in $\mathbf{q}$ are given by

$$
\begin{aligned}
d_0 &= \alpha_1\left(q_x+q_y\right)+\alpha_2\left(q_x+q_y\right)^2+\alpha_3\left(q_x-q_y\right)^2+\alpha_4\left(q_x^3+q_y^3\right)+\alpha_5\left(q_x^2q_y+q_y^2q_x\right),\\
d_x &= \beta_1\left(q_x-q_y\right)+\beta_2\left(q_x^2-q_y^2\right)+\beta_3\left(q_x^3-q_y^3\right)+\beta_4\left(q_x^2q_y-q_y^2q_x\right),\\
d_y &= 0,\\
d_z &= \gamma_1\left(q_x+q_y\right)+\gamma_2\left(q_x+q_y\right)^2+\gamma_3\left(q_x-q_y\right)^2+\gamma_4\left(q_x^3+q_y^3\right)+\gamma_5\left(q_x^2q_y+q_y^2q_x\right),
\end{aligned}
\tag{15}
$$

where $\alpha_i$, $\beta_i$ and $\gamma_i$ are the expansion coefficients. Within the linear **q**-order approximation, the Hamiltonian (13) yields Eq. (1) in the main text.